
\magnification=\magstep1
\baselineskip=20pt
\centerline{Yet Another Model of Soft Gamma Repeaters}
\bigskip
\centerline{J. I. Katz, H. A. Toole and S. H. Unruh}
\medskip
\centerline{Department of Physics and McDonnell Center for the Space
Sciences}
\centerline{Washington University, St. Louis, Mo. 63130}
\centerline{I: katz@wuphys.wustl.edu}
\bigskip
\centerline{Abstract}
\medskip
We develop a model of SGR in which a supernova leaves planets orbiting a
neutron star in intersecting eccentric orbits.  These planets will collide
in $\sim 10^4$ years if their orbits are coplanar.  Some fragments of debris
lose their angular momentum in the collision and fall onto the neutron star,
producing a SGR.  The initial accretion of matter left by the collision with
essentially no angular momentum may produce a superburst like that of March
5, 1979, while debris fragments which later lose their angular momentum
produce an irregular pattern of smaller bursts.
\vfil
\noindent
Subject headings: Accretion---Gamma-rays: Bursts---Stars: Neutron
\eject
\centerline{1. Introduction}
\medskip
More than a hundred models of gamma-ray bursts (GRB), including soft gamma
repeaters (SGR), have been published (Nemiroff 1994), yet there is at present
no satisfactory and generally accepted model of SGR.  The distinction
between ``classical'' GRB and SGR was not recognized until 1987 (Atteia,
{\it et al.}~1987; Laros, {\it et al.}~1987; Kouveliotou, {\it et al.}~1987),
so that theorists usually attempted to to explain both classes of
events with a single model.  However, most of the basic facts about SGR were
appreciated immediately following the observation of the superburst of March
5, 1979 (Cline 1980) from SGR 0526-66 in the LMC; specifically, their
distance scale, association with young supernova remnants, repetition and
likely origin in slowly rotating neutron stars were apparent.  Reported
annihilation spectral features were used to infer (Katz 1982) an upper
bound to the magnetic field of $\approx 10^{13}$ gauss, although recent
controversies regarding the reality of spectral features from ``classical''
GRB and the unclear spectrum of March 5, 1979 itself suggest that the
features and the bound on the field may be questionable (Duncan and
Thompson [1992] discuss models with much larger field).  More recently, a
distance scale of $\sim 10$ Kpc for the two non-LMC SGR (SGR 1826-20 and SGR
1900+14) and other evidence were reviewed by Norris, {\it et al.} (1991),
and at least two of the three SGR are now known to be associated with young
SNR (Kulkarni and Frail 1993; Kouveliotou, {\it et al.} 1994).

Two classes of models extensively discussed for GRB have appealing features
as explanations of SGR, but fail at least one crucial test.  Models
involving the accretion of comets, asteroids or planetesimals onto neutron
stars (Harwit and Salpeter 1973; Shklovskii 1974; Newman and Cox 1980;
Howard, Wilson and Barton 1981; Colgate and Petschek 1981; Van Buren 1981;
Joss and Rappaport 1984; Tremaine and \.Zytkow 1986; Katz 1986; Livio and
Taam 1987; Pineault and Poisson 1989; Pineault 1990; Colgate and Leonard
1994) are capable of explaining the energetics of SGR if they are at
distances $\sim 100$ pc (as classical GRB were long thought to be) but do
not release enough energy to explain them at the required distances across
the Galaxy or in the LMC, at which energies $\sim 10^{41}$ erg are required
($\sim 4 \times 10^{44}$ erg for March 5, 1979; Cline 1980).

Models involving processes resembling scaled-up Solar flares (Stecker and
Frost 1973; Katz 1982; Liang and Antiochos 1984; Melia and Fatuzzo 1991;
Katz 1993, 1994) have been popular because both flares and GRB have
irregular spiky time structure, while both flares and SGR repeat at
irregular intervals.  The energy of neutron star magnetospheres with $B <
10^{13}$ gauss is sufficient to explain all SGR phenomena except the
superburst of March 5, 1979 (where a deficiency $O(10)$ might
optimistically be attributed to anisotropic emission, a larger field, or to
regeneration by an interior rotational dynamo).  However, models of this
type are unable to explain any of the temporal or spectral properties of
SGR, their rarity, or their association with young supernova remnants, and
must attribute them to the mysterious properties of flares, whether on the
Sun, other stars, or neutron stars.  It is also difficult to
see how to produce a flare in a magnetosphere (where current paths are open
and force-free) or within a neutron star (where the conductivity and density
are extremely high, and any energy dissipated is thermalized and diffuses
slowly); see Carrigan and Katz (1992).  Further, models involving magnetic
flares or internal rotational relaxation might be expected to be associated
with pulsar glitches (Pacini and Ruderman 1974; Tsygan 1975), but no such
association has ever been observed.

An important clue is the rarity of SGR.  Kouveliotou, {\it et al.} (1994)
established that there are $< 7$ (and possibly only the known 3) SGR
presently active in our Galaxy, implying that either the SGR phase is only
$\le 10\%$ of the first $10^4$ years of a neutron star's life (studies of
the SNR in which SGR are embedded imply that the neutron stars which produce
SGR are no more than $\sim 10^4$ years old) or that only $\le 10\%$ of
neutron stars ever become SGR.  There is something special about these
neutron stars.

The model we suggest depends on the discovery (Wolszczan and Frail 1992) of
two planets orbiting PSR 1257+12.  Although this is a fast, low-field and
probably old (characteristic age $8 \times 10^8$ years) pulsar, we draw an
analogy with the young, slow (8 sec in SGR 0526-66) and probably high-field
(to produce the rotational modulation of March 5, 1979) neutron stars
associated with SGR.  The planets of PSR 1257+12 have been argued to be the
residue of the evaporation of a companion star, but it is unclear how to
form planets at a distance of $\approx 0.5$ AU from an evaporating companion
whose orbital radius (inferred from binary pulsars presently observed to be
evaporating their companions) is $\sim 10^{-2}$~AU.  Instead, we suggest
that the planets of PSR 1257+12 existed before and survived its formation,
and that similar surviving planets are the key to understanding SGR.

If planets survive a supernova which produces a neutron star, they will be
left in eccentric orbits as a result of loss of stellar mass and of
ablation.  The longitudes of periastron of these orbits depend on the
planets' longitudes at the time of the supernova, and will be uncorrelated.
If the orbits are coplanar, the eccentricities are substantial (as observed
for some binary pulsars, whose eccentricities are similarly produced by
supernova mass loss) and the initial semi-major axes were comparable (as
observed for terrestrial planets in the Solar system, and for the planets of
PSR 1257+12) then there is a substantial probability the orbits will
intersect and the planets will collide.

Both colliding planets have the same sense of angular momentum, but the
angular momentum of individual mass elements is not conserved in the
collision (its integral is conserved, of course), so that some of the
collision debris may have zero or very small angular momentum.  These
fragments will quickly fall onto the neutron star.  Fragments with somewhat
larger angular momentum may later lose it, producing an accretional rain
extending over a lengthy period as they are depleted.  A history qualitatively
resembling that of SGR 0526-66 is suggested, with a single superburst
followed by a long period of occasional smaller bursts.  If this is correct
(it is not required by the rest of the model) SGR 1806-20 and SGR 1900+14 had
superbursts which preceded the development of gamma-ray astronomy.

In \S2 we discuss the survival of planets orbiting a supernova.  \S3
contains an estimate of the rate of planetary collisions.  The problem of
angular momentum distribution and evolution of debris is briefly discussed
in \S4.  Expected and observed radiation properties are compared in \S5.
All of these discussions are preliminary and rough; more careful calculation
will be justified only if the model survives preliminary scrutiny.  \S6
contains a brief summary discussion.
\bigskip
\centerline{2. Planetary Survival}
\medskip
Planets must first survive the pre-supernova star.  Consider a planet
of mass $M_p = 3 M_E$ ($M_E$ is the Earth's mass), the minimum mass
permitted for the planets of PSR 1257+12, and iron composition, in circular
orbit of radius $R = 0.5$ AU (again resembling PSR 1257+12, although the
pre-supernova orbit may be somewhat smaller than a circularized
post-supernova orbit).  More massive or more distant planets will be more
robust, but the assumed parameters are sufficient to make survival likely.
It is probably necessary that the planet avoid engulfment by the
pre-supernova star, which would lead both to evaporation in the interior
radiation field and to a rapid inward death spiral resulting from
hydrodynamic drag.  The radii of pre-supernova stars are controversial, and
it is possible that engulfment may be avoided with $R = 0.5$ AU;
alternatively, it is not unlikely that some pre-supernovae will be
accompanied by planets (less readily detectable by pulsar timing studies)
with orbits large enough to avoid engulfment even by a red giant; the
existence of the major planets in the Solar system is evidence for the
existence of such planets.

The surface effective temperature $T_e$ of a planet, averaged over its
surface, is
$$T_e = \left({L \epsilon_a \over 16 \pi R^2 \epsilon_e \sigma_{SB}}
\right)^{1/4} \approx 5000^{\circ\ }{\rm K}, \eqno(1)$$
where $L$ is the stellar luminosity, $\epsilon_a$ its mean absorptivity,
$\epsilon_e$ its mean emissivity, and $\sigma_{SB}$ the Stefan-Boltzmann
constant; we have taken $\epsilon_a = \epsilon_e$ and $L = 10^{38}$ erg/sec,
appropriate to a luminous pre-supernova star.  The thermal energy should be
compared to the surface binding energy of an iron atom of mass $m_{\rm Fe}$:
$${k_B T_e a \over G M_p m_{\rm Fe}} \approx 0.005, \eqno(2)$$
where we have taken the planetary radius $a = 8 \times 10^8$ cm, the radius
of a zero-temperature iron planet of $3 M_E$ (Zapolsky and Salpeter 1969).
If the iron atmosphere is, on average, singly ionized, as appropriate at
$T_e$, its molecular weight is $m_{\rm Fe}/2$ and the value in Eq. (2) should
be doubled.  The Boltzmann factor for thermal evaporation is then $\sim
\exp(-100) \sim 10^{-43}$, and the rate of evaporation is negligible.

The thermal diffusivity of hot iron (using data for solid iron near its
melting temperature) is $\sim 0.1$ cm$^2$/sec.  In the $\sim 10^6$ year
lifetime of a bright pre-supernova star the surface heat diffuses only $\sim
2 \times 10^6$ cm into the interior; most of the interior remains cool and
solid.  In any case, the estimated $T_e$ would not make a large change in
the equation of state at the characteristic pressure $P_p \sim G M_p^2/a^4
\sim 5 \times 10^{13}$ erg/cm$^3$.  As a result, the zero-temperature value
of $a$ remains a good approximation.

The next threat to a planet is the supernova itself.  The near-circular
orbits observed for PSR 1257+12 may perhaps be explained by a quiet collapse
without mass loss other than neutrino radiation, but our SGR model requires
substantial eccentricity and mass loss.  We consider a low energy supernova
resembling the Crab supernova, whose debris shell has mass $\Delta M = 0.2
M_\odot$ and velocity $v_{SN} = 2 \times 10^8$ cm/sec, and a high energy
supernova with $\Delta M = M_\odot$ and $v_{SN} = 10^9$ cm/sec.  The impulse
delivered by the momentum in the debris shell corresponds to an impulsive
velocity change of the planet
$$\Delta v \approx {\Delta M v_{SN} a^2 \over 4 R^2 M_p} \approx 3 {\Delta M
\over M_\odot}{v_{SN} \over 10^9 {\rm cm/sec}} \ {\rm km/sec}, \eqno(3)$$
Even for the energetic supernova the delivered impulse is sufficient only to
induce an eccentricity $< 0.1$, and insufficient to contribute significantly
to disruption of the planetary orbit.

The intercepted kinetic energy $KE$ is more important, and its ratio to
the planet's characteristic binding energy $E_{bind} \sim G M_p^2/a$ is
$${KE \over E_{bind}} \sim {\Delta M v_{SN}^2 a^3 \over 8 R^2 G M_p} \sim
100 {\Delta M \over M_\odot} \left({v_{SN} \over 10^9 {\rm
cm/sec}}\right)^2, \eqno(4)$$
For a Crab-like supernova the intercepted kinetic energy is insufficient to
disrupt the planet, and a numerical hydrodynamic calculation would
probably show that most of it is concentrated in surface blowoff and
radiation, leaving the planet's dense core unscathed.  However, the more
energetic supernova would imply $KE \sim 100 E_{bind}$ and requires more
careful attention.

In fact, a supernova debris shell is spread by adiabatic expansion into a
wind whose thickness $\Delta R$ is $\sim 0.3 R$ at a distance $R$.  Its
stagnation pressure $P_{SN}$ is
$$P_{SN} \sim {\Delta M v_{SN}^2 \over 8 \pi R^2 \Delta R} \sim 10^{12}
{\Delta M \over M_\odot} \left({v_{SN} \over 10^9 {\rm cm/sec}}\right)^2\
{\rm erg/cm}^3. \eqno(5)$$
Even for the energetic supernova, $P_{SN} \ll P_p$, so that the pressure of
impacting debris is insufficient to disrupt the planet, although there will
be some (difficult to calculate, even numerically) mass loss by ablation and
surface entrainment.  The kinetic energy of the supernova wind flows around
the planet, guided by the stagnation pressure at the planetary surface.  It
is not coupled into the planetary interior and does not disrupt the planet.
A useful (though inexact) hydrodynamic analogy is a spacecraft re-entering
the Earth's atmosphere, whose kinetic energy far exceeds its heat of
vaporization, but which suffers very little ablation.

Ablation will produce a recoil which depends on the amount of energy
hydrodynamically coupled to the planetary interior.  This is also difficult
to calculate.  Because the ablation pressure is small (even for energetic
supernovae) and the density mismatch is large between the wind, which has
density $\sim \Delta M/(4 \pi R^2 \Delta R) \sim 10^{-6}$ gm/cm$^3$ and the
planet ($\sim 10$ gm/cm$^3$), this coupling is likely to be poor, and the
resulting velocity of planetary recoil will be of order that given by Eq.
(3).

It thus appears that a planet with the assumed parameters would survive the
pre-supernova star and a Crab-like supernova essentially intact, and would
probably also survive even the most energetic supernovae.  Its orbit is,
however, affected by the loss of mass in the supernova explosion, according
to the classic theory of Blaauw (1961).  If less than half the
pre-supernova's mass is lost in a symmetric (recoiless) explosion then the
planet will remain bound, but in an eccentric orbit.

Recoil by the newly formed neutron star would, itself, unbind the planet if
the recoil velocity exceeded the planetary orbital velocity (about 50 km/sec
for the assumed parameters).  Most pulsars have velocities substantially
exceeding this value.  However, there is evidence (Katz 1975) from the
presence of neutron stars in globular clusters for their production with
recoil velocities less than the cluster escape velocities, which are
typically $\sim 20$ km/sec.

Kulkarni, {\it et al.} (1994) and Rothschild, Kulkarni and Lingenfelter
(1994) have argued that the neutron stars producing SGR 1806-20 and SGR
0526-66 are offset from the centers of their supernova remnants by
amounts corresponding to space velocities of 500 km/sec and 1200 km/sec,
respectively, inconsistent with the hypothesis of this paper.  The
brightness distributions of these remnants are rather irregular; if the
observed shapes of the remnants are attributed to the collision of supernova
debris with asymmetrically located interstellar clouds then it may be that
the SGR are actually in the dynamical (rather than the X-ray or radio)
centers of their supernova remnants, reconciling these observations with the
low recoil velocities required by our hypothesis.

The planets orbiting PSR 1257+12 have small eccentricities ($e \approx
0.02$).  Thus, while they may be taken as evidence for the survival of
planets (though post-supernova formation has been considered), they could be
considered arguments against significant eccentricity of the planetary
orbits.  However, the characteristic (spin-down) age of PSR 1257+12 is $8
\times 10^8$ years, and its actual age could be of this order, affording
ample time for weak dissipative processes (such as interaction with the
pulsar's radiation or a residual disk) to circularize the planetary orbits.
In addition, PSR 1257+12 has a spin period of 6 ms and a correspondingly
intense radiation field, in contrast to the 8 sec period inferred for SGR
0526-66, so these systems may differ in many aspects of their history and
properties.
\bigskip
\centerline{3. Planetary Collision Rate}
\medskip
We assume intersecting coplanar eccentric planetary orbits with semi-major
axes $R \approx 10^{13}$ cm.  The orbital velocities ($\approx$ 50 km/sec)
exceed the escape velocities from the planets ($\approx$ 20 km/sec) so that
their collision cross-sections are essentially geometrical.  In each orbit
of circumference $\approx 2 \pi R$ the length which corresponds to a
planetary collision is $8a$, allowing for two intersections (any approach of
the planetary centers to within a separation of $2a$ in any direction leads
to collision) and ignoring the fact that the angle between the orbits at
intersection is likely to be substantially less than $\pi/2$.  The
probability of collision per orbit is $\approx 8a/(2 \pi R)$, and the
characteristic time to collision $t_c$ is
$$t_c \approx P {2 \pi R \over 8 a} \approx 4000\ {\rm years}, \eqno(6)$$
where we have taken the synodic period of one planet with respect to the
other $P \approx 0.5$ year, following PSR 1257+12.  This value of $t_c$ is
consistent with the observational inference that SGR occur in supernova
remnants which are no more than $10^4$ years old.

The assumption of coplanar orbits is essential to Eq.~(6); the relative
inclination angle $i$ must not greatly exceed $2a/R \sim 2 \times 10^{-4}$
rad.  This is substantially smaller than inclinations found in our Solar
system, in which relative inclinations are $\sim 1^\circ$ or more.
However, the pre-supernova star is very extended, and planets will have
strong tidal interactions with it, and through it with each other.  It is
therefore not implausible that their orbits will relax to accurate
coplanarity.  The closest Solar system analogs, the satellite systems of the
major planets, although closer to coplanarity than the Solar system itself,
are not directly applicable because the major planets have large equatorial
bulges (not expected for a giant star) and their satellites are also
perturbed out of their orbital planes by other planets and the Sun.

If the orbits are not coplanar, with $i \gg 2a/R$, the planets will not
collide unless there is a fortuitous coincidence of a node with one of the
longitudes of equal radii.  Differential precession and apsidal motion may
produce approximate intersection and collision, but $t_c$ is increased by a
factor $O(R \sin i/2a)$ if the motion is ergodic; there is also the
possibility of a stable non-colliding state such as that between Neptune
and Pluto.
\bigskip
\centerline{4. Collision Debris}
\medskip
Two colliding planets will both have the same sense of angular momentum.
How may they then produce debris in orbits of essentially zero angular
momentum, as required for accretion onto the neutron star?  In a collision
there will be large internal forces which will redistribute momentum and
angular momentum over the colliding planets.  Materials strength and
gravitational binding are small compared to the typical collisional stresses
of $\sim 10^{14}$ erg/cm$^3$ for planets whose orbital velocities are about
50 km/sec, so much or all of the planets (depending on impact parameter)
will be disrupted into a spray of fragments of various sizes.  The specific
angular momentum distribution is unknown, so we will assume it to be uniform
from $-0.5 L_0$ to $2.5 L_0$, where $L_0$ is the mean initial specific
angular momentum.  The results depend on the value of the distribution at $L
\approx 0$ (it is essential that it be nonzero there) but not on its shape;
the assumption of some debris with $L < 0$ is relevant to subsequent
debris-debris collisions.

The accretion of an asteroid by a magnetic neutron star was discussed by
Colgate and Petschek (1981); similar processes occur here, but the
greater mass of the accreting object makes the magnetic field less important
in the dynamics.  Debris with low angular momentum moves in essentially
parabolic orbits, with periastron distance $h = L^2/(2GM)$.  A neutron star
magnetic moment $\mu$ will stop an iron fragment of $s = 3 \times 10^6$ cm
radius (required to explain a typical SGR burst of $10^{41}$ erg) at a
periastron distance
$$h_s < \left({\mu^2 \over 8 \pi G M \rho s}\right)^{1/4} \approx 1.7 \times
10^6 \left({\mu \over 10^{30}\ {\rm gauss\,cm}^3}\right)^{1/2} \left({8\
{\rm gm/cm}^3 \over \rho}\right)^{1/4}\ {\rm cm}. \eqno(7)$$
This will not exceed the neutron star's radius in order of magnitude.
Compression of the infalling matter by the neutron star's gravitational
field only strengthens this conclusion.

The limit on $L$ for accretion by direct infall from the collision site is
approximately that required for ballistic impact with the surface of a
neutron star of radius $r$:
$$L < (2GMr)^{1/2} \approx 2 \times 10^{16}\ {\rm cm^2/sec}. \eqno(8)$$
For orbits like those of the planets of PSR 1257+12 $L_0 \approx 4 \times
10^{19}$ cm$^2$/sec.  The total mass available for prompt infall is
$$M_{sb} \sim 2 M_p {L \over 3 L_0} \sim 6 \times 10^{24}\ {\rm gm}.
\eqno(9)$$
The gravitational energy released when this mass accretes is $\sim 6 \times
10^{44}$ erg, approximately that required (Cline 1980) to explain the
superburst of March 5, 1979.

This prompt accretion occurs roughly 0.2 of an orbital period following the
collision, or some weeks later.  In order to explain the observed
superburst of March 4, 1979, which had a very intense sub-burst lasting
$\sim 0.1$ second followed by most of its energy over a period of several
minutes, the initial accretion must have been of a single solid body of
$\sim 10^{25}$ gm, rather than an extended cloud of small particles or fluid.
This is not impossible; the planetary interiors are expected to be solid
metal (\S2).

Subsequent repeating bursts require the accretion of smaller bodies,
typically $\sim 10^{21}$ gm but with wide dispersion.  A few of these may
have left the collision with $L$ small enough to collide with the neutron
star's surface, but with more outward-directed velocity than the fragment
accreted first, so their accretion follows the superburst by weeks to
months.  To explain repetitions at longer times requires fragments born with
greater $L$ which they lose over a longer period.  A number of mechanisms
are possible, including angular momentum loss to the magnetosphere in
non-impacting periastron passages (Van Buren 1981), and subsequent
collisions between debris or between debris and surviving planetary cores
(possibly including other planets not involved in the collision).

Most of these processes are difficult to calculate.  The resulting temporal
distribution of bursts depends on the detailed angular momentum
distribution of the collision debris, which is also unknown.  Stochastic
gravitational perturbation of debris by planets may be estimated, and amounts
to $\sim 2 \times 10^{14}$ cm$^2$/sec per orbit.  This suggests that in
$\sim 10^4$ orbits (several thousand years) the total amount accreted might
be comparable to that in the initial superburst, roughly consistent with
observation.

The tidal breakup radius of an iron fragment of radius $s = 3 \times 10^6$ cm
($\sim 10^{21}$ gm) and strength $Y \sim 10^{10}$ erg/cm$^3$ is $r_t \sim
(GM\rho s^2 /2Y)^{1/3} \sim 10^{10}$ cm (the nominal Roche limit is several
times larger, but irrelevant in the presence of this material strength),
which poses a problem for all gradual orbital evolution processes---the
fragment will be disrupted before it loses enough angular momentum to be
accreted.  Disruption produces a mix of smaller fragments and perhaps fluid,
which may accrete steadily through a disk (if not blown away by the pulsar
wind), leading to a steady X-ray source of low luminosity.  Debris-debris
collisions or rare close encounters with planets may resolve this problem by
impulsively transferring fragments with $L > (2GMr_t)^{1/2}$ into collision
orbits with the neutron star, without requiring them to survive diffusion
through values of angular momentum at which they would be disrupted without
accretion.
\bigskip
\centerline{5. Radiation}
\medskip
The observed spectra of SGR are typically fitted to optically thin
bremsstrahlung spectra with $k_B T \approx 40$ KeV, but may perhaps also be
fitted to black bodies with $k_B T \approx 10$--20 KeV.  Such a black body
with a neutron star's surface as its radiating area is consistent with the
luminosities $\sim 10^{41}$ erg/sec typically observed, as may be the much
harder spectrum observed during the more luminous initial 0.1 second of the
March 5, 1979 superburst.

The most striking fact about these luminosities, as pointed out by Cline
(1980) for SGR 0526-66 and by Atteia, {\it et al.} (1987) and Kouveliotou,
{\it et al.} (1994) for the other SGR, is that they are apparently
super-Eddington by several orders of magnitude.  Two explanations are
possible.  One is that they are not really super-Eddington, because the
neutron stars' magnetic fields are sufficiently large that the opacity at
the observed frequencies is far below that of free electron scattering for
radiation propagating with its electric vector perpendicular to ${\vec B}$.
This open spectral/polarization ``window'' dominates the Rosseland mean
opacity, and increases the Eddington limit above its nominal (unmagnetized
electron scattering) value.  This hypothesis requires magnetic fields in
excess of $10^{13}$ gauss, and predicts strong linear polarization of the
emergent radiation.

In the second explanation the luminosity is genuinely super-Eddington, but
the radiation pressure is contained by the magnetic stress (Katz 1982),
rather than the the weight of overlying matter.  In this case the radiation
energy which may be contained is limited to approximately the magnetic
energy of the magnetosphere.  For $B < 10^{13}$ gauss this is adequate to
explain the ordinary bursts of SGR but not the superburst of March 5, 1979.
\bigskip
\centerline{6. Discussion}
\medskip
We have presented a model for SGR which may explain, at least in order of
magnitude, many of their properties.  The model makes certain testable
predictions.  A SGR may (but need not, if no fragments are born satisfying
Eq. [8]) begin its activity with a superburst like that of March 5, 1979;
however, if such a superburst occurs it will not have been preceded by
years of ordinary bursts.  The radiation is predicted to be linearly
polarized, with approximately a black body spectrum.  The magnetic field of
the neutron star probably exceeds $10^{13}$ gauss, in which case no 511 KeV
annihilation line can be seen (Katz 1982).  A SGR has negligible visible or
radio-frequency emission.  Following a burst, the luminosity should decay
roughly $\propto t^{-3/2}$, the result for the cooling of an impulsively
heated half-space of uniform thermal impedance (Katz 1982); more
quantitative results require numerical calculation with realistic neutron
star models.

It is possible that in the superburst of March 5, 1979 the initial intense
$\sim 0.1$ second of emission reflected the duration of rapid accretion,
while the subsequent emission resulted from the cooling of accreted matter.
If so, then the duration of accretion reflects the tidal breakup of the
accreting body on infall.  An elementary calculation yields a duration $4 s
(r_\ell/2GM)^{1/2}/5$, where $r_\ell \equiv \min (r_t, r_{Roche})$; for $s =
5 \times 10^7$ cm, corresponding to the total energy of the superburst,
this is 0.4 sec, almost as short as required.

The submillisecond rise time (Cline 1980) of the March 5, 1979 superburst
requires explanation.  When the compressed iron of a disrupted fragment hits
the neutron star it produces a splash of radiating shock-heated
matter, which is ballistically distributed over the surface on the
submillisecond gravitational time scale (Howard, Wilson and Barton 1981).
The rise time is short because the accreting matter is condensed, rather
than gaseous; the rate of release of hydrodynamic (infall) energy rises
abruptly from zero just as the density of solid iron rises abruptly from
vacuum.  The duration of energy release is much longer, and is set by the
duration of infall of the disrupted accreting fragment.

The collision of two planets releases $\sim 10^{41}$ erg, but nearly all of
this energy appears in thermal and kinetic energy of debris, and is not
radiated.  The luminosity of the shock-heated planets themselves resembles
that of a hot white dwarf ($M_V \sim 10$), which at 10 Kpc distance
corresponds to $m_V \sim 25$, surely undetectable as a rare and
unpredictable event preceding gamma-ray emission.  The spray of debris
could have a surface area larger by orders of magnitude.  In the most
extreme and implausibly optimistic case most of the planets could be
vaporized, and their effective radiating area could be $\sim R^2$ after a
time of order the orbital period.  The kinetic energy, if all radiated from
this area in this time, would give $M_V \sim 4$ and $m_V \sim 19$.

We thank P. C. Joss and S. R. Kulkarni for discussions and NASA NAGW-2918
for support.  S. H. Unruh extends special thanks to H. A. D. Gadya.
\bigskip
\centerline{References}
\parindent=0pt
\def\ref{\medskip \hangindent=20pt \hangafter=1}
\ref
Atteia, J.-L., {\it et al.} 1987 ApJ 320, L105
\ref
Blaauw, A. 1961 Bull. Astr. Inst. Neth. 15, 265
\ref
Carrigan, B. J., and Katz, J. I. 1992 ApJ 399, 100
\ref
Cline, T. 1980 Comments on Ap. 9, 13
\ref
Colgate, S. A., and Leonard, P. J. T. 1994 in Huntsville Gamma-Ray Burst
Workshop eds. G. Fishman, K. Hurley, J. Brainerd (AIP, New York) in press
\ref
Colgate, S. A., and Petschek, A. G. 1981 ApJ 248, 771
\ref
Duncan, R. C., and Thompson, C. 1992 ApJ 392, L9
\ref
Harwit, M., and Salpeter, E. E. 1973 ApJ 186, L37
\ref
Howard, W. M., Wilson, J. R., and Barton, R. T. 1981 ApJ 249, 302
\ref
Joss, P. C., and Rappaport, S. A. 1984 in High Energy Transients in
Astrophysics ed. S. E. Woosley (AIP, New York) 555
\ref
Katz, J. I. 1975 Nature 253, 698
\ref
Katz, J. I. 1982 ApJ 260, 371
\ref
Katz, J. I. 1986 ApJ 309, 253
\ref
Katz, J. I. 1993 in Compton Gamma-Ray Observatory eds. M. Friedlander, N.
Gehrels, D. J. Macomb (AIP, New York) 1090
\ref
Katz, J. I. 1994 ApJ 422, 248
\ref
Kouveliotou, C., {\it et al.} 1987 ApJ 322, L21
\ref
Kouveliotou, C., {\it et al.} 1994 Nature 368, 125
\ref
Kulkarni, S. R., {\it et al.} 1994 Nature 368, 129
\ref
Kulkarni, S. R., and Frail, D. A. 1993 Nature 365, 33
\ref
Laros, J. P., {\it et al.} 1987 ApJ 320, L111
\ref
Liang, E. P. and Antiochos, S. K. 1984 Nature 310, 121
\ref
Livio, M., and Taam, R. E. 1987 Nature 327, 398
\ref
Melia, F., and Fatuzzo, M. 1991 ApJ 373, 198
\ref
Nemiroff, R. J. 1994 Comments on Ap. in press
\ref
Newman, M. J., and Cox, A. N. 1980 ApJ 242, 319
\ref
Norris, J. P., Hertz, P., Wood, K. S., and Kouveliotou, C. 1991 ApJ 366,
240
\ref
Pacini, F., and Ruderman, M. 1974 Nature 251, 399
\ref
Pineault, S. 1990 Nature 345, 233
\ref
Pineault, S., and Poisson, E. 1989 ApJ 347, 1141
\ref
Rothschild, R. E., Kulkarni, S. R., and Lingenfelter, R. E. 1994 Nature 368,
432
\ref
Shklovskii, I. S. 1974 Sov. Astr. 18, 390
\ref
Stecker, F. W., and Frost, K. J. 1973 Nature Phys. Sci. 245, 70
\ref
Tremaine, S., and \.Zytkow, A. N. 1986 ApJ 301, 155
\ref
Tsygan, A. I. 1975 Astron. Ap. 44, 21
\ref
Van Buren, D. 1981 ApJ 249, 297
\ref
Wolszczan, A., and Frail, D. A. 1992 Nature 355, 145
\ref
Zapolsky, H. S., and Salpeter, E. E. 1969 ApJ 158, 809
\vfil
\eject
\bye
\end